\documentclass{article}
\usepackage{amssymb}

\usepackage{graphicx}
\usepackage{amsmath}

\begin{document}

\title{Phenomenology of lepton flavor violation in 2HDM(III) from\ $\left(
g-2\right) _{\mu }\;$and leptonic decays.}
\author{Rodolfo A. Diaz, R. Martinez, and J.-Alexis Rodriguez \\
Universidad Nacional de Colombia.\\
Departamento de Fisica. Bogota, Colombia}
\maketitle

\begin{abstract}
We constrain some lepton flavor violating vertices in the context of the two
Higgs doublet model type III, assuming that the lightest scalar Higgs mass $%
m_{h^{0}}\;$is about $115\;$GeV. Specifically, based on the $g-2\;$muon
factor and the decay width of$\;\mu \rightarrow e\gamma $,$\;$the following
quite general bounds are obtained:$\;7.62\times 10^{-4}\lesssim \xi _{\mu
\tau }^{2}\lesssim 4.44\times 10^{-2},\;$ $\xi _{e\tau }^{2}\lesssim
2.77\times 10^{-14}$. Additionally, based on the processes $\tau \rightarrow
\mu \gamma $, and $\tau \rightarrow \mu \mu \mu $, bounds on $\xi _{\tau
\tau }\;$and $\xi _{\mu \mu }\;$are also gotten, such constraints on these
parameters$\;$still give enough room for either a strong suppression$\;$or
strong enhancement on the coupling of any Higgs boson to a pair of tau
leptons or a pair of muons\textbf{.}\ Furthermore, upper limits on the decay
widths of the leptonic decays $\tau \rightarrow e\gamma ,\;$and $\tau
\rightarrow eee\;$are calculated, finding them to stay far from the reach of
near future experiments.
\end{abstract}

\section{Introduction}

It is well known that processes involving Flavor Changing Neutral Currents
(FCNC), are severely suppressed by experimental data, despite they seem not
to violate any fundamental law of nature. On the other hand, Standard Model
(SM) issues are compatible with experimental constraints on FCNC so far,
with the remarkable exception of neutrino oscillations. Preliminary
measurements of the flux of solar neutrinos \cite{Davis} yielded a result
considerably smaller than the one expected from the standard solar model,
from which the possibility of oscillation of neutrinos arose as a possible
solution to the neutrino deficit observed \cite{Pontecorvo}. Later on,
neutrino oscillations in matter were proposed to explain the deficit
confirmed by SuperKamiokande \cite{Hirata}. Additional experiments with
solar \cite{Kameda} and atmospheric neutrinos \cite{Fukuda} have supported
the idea of the oscillations.

On the other hand, this phenomenon implies the existence of mass terms for
neutrinos, and the existence of family lepton flavor violation. Both
implications lead us in turn to consider the existence of physics beyond the
SM, since in the SM neutrinos are massless and lepton flavor violating
mechanisms are absent. Therefore, the increasing evidence on neutrino
oscillations motivates the study of models with lepton flavor violation.
There are some interesting scenarios in which FCNC and therefore LFV appears
naturally. For example, the introduction of new representations of fermions
different from doublets produce FCNC by means of the Z-coupling \cite{2}. On
the other hand, they could arise at the tree level in the Yukawa sector if a
second doublet is added to the SM \cite{wolf}. Another interesting scenario
appears in SUSY theories with R-parity broken \cite{R1}, because FCNC coming
from R-parity violation generates massive neutrinos and neutrino
oscillations \cite{Kaustubh}. FCNC in SUSY theories in the charged lepton
sector have been studied as well. For instance, the decays $\mu \rightarrow
e\gamma \;$and $\mu \rightarrow 3e\;$with polarized muons have been examined
in the context of supersymmetric grand unified theories \cite{Okada}.

Moreover, neutrino oscillations have inspired the study of Lepton Flavor
Violating (LFV) processes in the neutral leptonic sector. For example, LFV
processes in SU(5) SUSY models with right handed neutrinos have been
examined based on recent results of neutrino oscillation experiments \cite
{Baek}. On the other hand, ref \cite{Lavoura} explores the generation of LFV
by using a multi-Higgs doublet model with additional right handed neutrinos
for each lepton generation, finding a non-decoupling behavior of some LFV
amplitudes respect to the Right handed neutrino masses.

Additionally, experimental search has imposed some upper limits for the
branching of many processes involving LFV as in the case of $%
K_{L}^{0}\rightarrow \mu ^{+}e^{-}$ \cite{arisaka}, $K_{L}^{0}\rightarrow
\pi ^{0}\mu ^{+}e^{-}$ \cite{plb}, $K^{+}\rightarrow \pi ^{+}\mu ^{+}e^{-}$ 
\cite{lee}, $\mu ^{+}\rightarrow e^{+}\gamma $ \cite{bolton}, $\mu
^{+}\rightarrow e^{+}e^{+}e^{-}$ \cite{bell} and $\mu ^{-}N\rightarrow e^{-}N
$ \cite{doh}. As for experimental prospects, muon colliders offer a very
appealing alternative because of its potential to test the mixing between
the second and third (first) generations by processes like $\mu \mu
\rightarrow \mu \tau (e\tau )\;$ mediated by Higgs exchange \cite{Reina}, 
\cite{SherCollider},\cite{Workshop}.

Perhaps the simplest framework to look for these rare processes is the Two
Higgs Doublet Model (2HDM), which consists of considering the possibility of
duplication of the SM Higgs doublet. Owing to the addition of the second
doublet, the Yukawa interactions lead naturally to tree level FCNC unless we
make additional assumptions. The most common one is the imposition of a
discrete symmetry \cite{gw} which in turn generates two independent sets of
Yukawa interactions known as models type I and II. Notwithstanding, we can
still have the possibility of preserving all terms in the Yukawa sector
(known as model Type III) and obtain constraints on the matrix elements by
experimental data. For example, FCNC at the tree level could be compatible
with experiments in certain regions of the parameter space if we consider
the exchange of heavy scalar or pseudoscalar Higgs fields \cite{Sher91} or
by cancellation of large contributions with opposite sign.

Finally, a very important test of FC vertices in leptons could be provided
by the $\left( g-2\right) /2\;$muon factor $\left( a_{\mu }\right) $. Recent
measurements of $a_{\mu }\;$at BNL \cite{g2exp} have reduced the uncertainty
by a factor of two from previous measurements. On the other hand, SM
estimations on $a_{\mu }\;$come from QED, electroweak, and hadronic
contributions \cite{Marciano}; the hadronic ones carry the bulk of the
uncertainties. In particular, the light by light hadron contributions have
been recently corrected \cite{g2th}. From the experimental and SM values of $%
a_{\mu }\;$it is possible to estimate the room for new physics $\left(
\Delta a_{\mu }\right) $ available for this observable. Several works
constraining LFV processes from previous estimations on $\Delta a_{\mu }\;$%
have been carried out \cite{g2us, KangLee}. In order to constrain such kind
of processes, we shall use the following interval for $\Delta a_{\mu }\;$at
95\% confidence level (C.L.), reported by \cite{Maria} 
\begin{equation}
9.38\times 10^{-10}\leq \Delta a_{\mu }\leq 51.28\times 10^{-10}
\label{Maria K}
\end{equation}

In this paper we examine constraints on the Flavor Changing (FC) vertices in
the lepton Yukawa sector of the 2HDM type III. Such bounds will be gotten
from $\Delta a_{\mu }$ and the leptonic decays $\mu \rightarrow e\gamma
,\;\tau \rightarrow \mu \gamma \;$and $\tau \rightarrow \mu \mu \mu $.$\;$In
addition, we predict upper limits for the decays $\tau \rightarrow e\gamma \;
$and $\tau \rightarrow eee.$

\section{LFV Processes}

The Yukawa interactions in the 2HDM type III, relevant for our purposes are 
\begin{eqnarray}
-\pounds _{Y} &=&\frac{g}{2M_{W}}\overline{E}M_{E}E\left( \cos \alpha
H^{0}-\sin \alpha h^{0}\right) +\frac{1}{\sqrt{2}}\overline{E}\xi
_{E}E\left( \sin \alpha H^{0}+\cos \alpha h^{0}\right)  \notag \\
&+&\frac{i}{\sqrt{2}}\overline{E}\xi _{E}\gamma _{5}EA^{0}+\text{h.c.}
\label{Yuk}
\end{eqnarray}
where $H^{0}$($h^{0}$)$\;$denote the heaviest (lightest) neutral $CP-$even
scalar, and$\;A^{0}\;$is a $CP-$odd scalar. $E\;$refers to the three charged
leptons $E\equiv \left( e,\mu ,\tau \right) ^{T}$ \ and $M_{E},\;\xi _{E}\;$%
are the mass matrix and the matrix of flavor changing vertices respectively.
Finally, $\alpha \;$is the mixing angle in the $CP-$even sector. We are
working in the simplest parametrization of the 2HDM type III in which only
one of the doublets acquire a Vacuum Expectation Value (VEV) \cite{Sher91}.

Now, we write down the expressions for the processes relevant in our
analysis. For $\Delta a_{\mu }$\ we shall use the integral expression \cite
{hunter} for the sake of accuracy. The expressions for $\Delta a_{\mu }$,
are given by

\begin{eqnarray}
\Delta a_{\mu } &=&\sum_{S}I_{S}+\sum_{P}I_{P}\;,  \notag \\
I_{S\left( P\right) } &=&C_{S\left( P\right) }^{2}\frac{m_{\mu }^{2}}{8\pi
^{2}}\int_{0}^{1}\frac{x^{2}\left( 1-x\pm m_{\tau }/m_{\mu }\right) }{m_{\mu
}^{2}x^{2}+\left( m_{\tau }^{2}-m_{\mu }^{2}\right) x+M_{S\left( P\right)
}^{2}\left( 1-x\right) }dx  \label{aNP}
\end{eqnarray}
where $I_{S\left( P\right) }\;$is an integral involving an Scalar
(Pseudoscalar) Higgs boson with mass $M_{S\left( P\right) }$,\ and $%
C_{S\left( P\right) }\;$is the corresponding coefficient in the Yukawa
Lagrangian (\ref{Yuk}). On the other hand, the expressions for $\Gamma
\left( \tau \rightarrow \mu \gamma \right) $, $\Gamma \left( \tau
\rightarrow e\gamma \right) $ and $\Gamma \left( \mu \rightarrow e\gamma
\right) \;$yield 
\begin{eqnarray}
\Gamma \left( \tau \rightarrow l\gamma \right)  &=&\xi _{l\tau }^{2}\frac{%
G_{F}\alpha _{em}m_{\tau }^{5}}{4\pi ^{4}\sqrt{2}}R\left(
m_{H^{0}},m_{h^{0}},m_{A^{0}},\alpha ,\xi _{\tau \tau }\right) \;,  \notag \\
\Gamma \left( \mu \rightarrow e\gamma \right)  &=&\xi _{\mu \tau }^{2}\xi
_{e\tau }^{2}\frac{\alpha _{em}m_{\tau }^{4}m_{\mu }}{16\pi ^{4}}S\left(
m_{H^{0}},m_{h^{0}},m_{A^{0}},\alpha ,\xi _{\tau \tau }\right) \;.
\label{tamufot}
\end{eqnarray}
$\;$ where $l\equiv e,\mu \;$denotes a light charged lepton. In addition, we
have defined 
\begin{eqnarray}
R\left( m_{H^{0}},m_{h^{0}},m_{A^{0}},\alpha ,\xi _{\tau \tau }\right) 
&=&\left| \left( m_{\tau }\sin 2\alpha +\frac{\sqrt[4]{2}\xi _{\tau \tau
}\sin ^{2}\alpha }{\sqrt{G_{F}}}\right) \frac{\ln \left[ m_{H^{0}}/m_{\tau }%
\right] }{m_{H^{0}}^{2}}\right.   \notag \\
&-&\left. \left( m_{\tau }\sin 2\alpha -\frac{\sqrt[4]{2}\xi _{\tau \tau
}\cos ^{2}\alpha }{\sqrt{G_{F}}}\right) \frac{\ln \left[ m_{h^{0}}/m_{\tau }%
\right] }{m_{h^{0}}^{2}}\right.   \notag \\
&-&\left. \frac{2\sqrt[4]{2}\xi _{\tau \tau }}{\sqrt{G_{F}}}\frac{\ln \left[
m_{A^{0}}/m_{\tau }\right] }{m_{A^{0}}^{2}}\right| ^{2}\;,  \notag \\
S\left( m_{H^{0}},m_{h^{0}},m_{A^{0}},\alpha ,\xi _{\tau \tau }\right) 
&=&\left| \sin ^{2}\alpha \frac{\ln \left[ m_{H^{0}}/m_{\tau }\right] }{%
m_{H^{0}}^{2}}+\cos ^{2}\alpha \frac{\ln \left[ m_{h^{0}}/m_{\tau }\right] }{%
m_{h^{0}}^{2}}\right.   \notag \\
&+&\left. \frac{\ln \left[ m_{A^{0}}/m_{\tau }\right] }{m_{A^{0}}^{2}}%
\right| ^{2}\;.  \label{R}
\end{eqnarray}
Finally, the expression for a lepton $L\;$going to three leptons of the same
flavor $l\;$is given by 
\begin{eqnarray}
\Gamma \left( L\rightarrow \overline{l}ll\right)  &=&\frac{m_{L}^{5}}{%
2048\pi ^{3}}\left[ \sin 2\alpha \sqrt{\frac{G_{F}}{\sqrt{2}}}\left( \frac{1%
}{m_{H^{0}}^{2}}-\frac{1}{m_{h^{0}}^{2}}\right) m_{l}\right.   \notag \\
&+&\left. \xi _{ll}\left( \frac{\sin ^{2}\alpha }{m_{H^{0}}^{2}}+\frac{\cos
^{2}\alpha }{m_{h^{0}}^{2}}-\frac{1}{m_{A^{0}}^{2}}\right) \right] ^{2}\;.
\label{Llll}
\end{eqnarray}

\subsection{Obtaining the bounds}

>From the previous section, the free parameters that we are involved with,
are: the three neutral Higgs boson masses $\left(
m_{h^{0}},m_{H^{0}},m_{A^{0}}\right) $,\ the mixing angle $\alpha $,$\;$and
some flavor changing vertices $\xi _{ij}$. Based on the present bounds from
LEP2 we shall assume that $m_{h^{0}}\approx 115\;$GeV. In addition, we shall
assume that $m_{A^{0}}\gtrsim m_{h^{0}}$, both assumptions will be held
throughout the document. Now, since we are going to consider plots in the $%
\xi _{ij}-m_{A^{0}}\;$plane, we should manage to use appropiate values of $%
\left( m_{H^{0}},\alpha \right) \;$in order to sweep a wide region of
parameters. In order to sweep a reasonable set of this couple of parameters
we utilize for $m_{H^{0}}\;$values of the order of $115\;$GeV (light), $300\;
$GeV (intermediate), and very large masses (heavy). As for the angle $\alpha 
$,\ we consider values of $\alpha =0\;$(minimal mixing), $\alpha =\pi /4\;$%
(intermediate mixing), and $\alpha =\pi /2$\ (maximal mixing). We can check
that all possible combinations of $\left( m_{H^{0}},\alpha \right) $ could
be done by considering five cases 1) when $m_{H^{0}}\simeq 115\;$GeV;$\;$ 2)
when $m_{H^{0}}\simeq 300\;$GeV$\;$and $\alpha =\pi /2;\;$ 3) when $%
m_{H^{0}}\;$is very large and $\alpha =\pi /2$; 4) when $m_{H^{0}}\simeq
300\;$GeV$\;$and $\alpha =\pi /4$;$\;$5) when $m_{H^{0}}\;$is\ very\ large,
and $\alpha =\pi /4$.$\;$In all these cases the mass of the pseudoscalar
will be considered in the range $115\;$GeV$\lesssim m_{A^{0}}$.

\begin{table}[tbp]
\begin{center}
\centering
\vskip0.1 in 
\begin{tabular}{||l||l|l|l||}
\hline\hline
& Bounds$\;$on$\;\xi _{\mu \tau }^{2}$ & Bounds$\;$on$\;\xi _{e\tau }^{2}\xi
_{\mu \tau }^{2}$ & Bounds$\;$on$\;\xi _{e\tau }^{2}$ \\ \hline\hline
case 1 & $7.62\times 10^{-4}\lesssim \xi _{\mu \tau }^{2}\lesssim
\allowbreak 8.31\times 10^{-3}$ & $\xi _{e\tau }^{2}\xi _{\mu \tau
}^{2}\lesssim 7.33\times 10^{-18}$ & $\xi _{e\tau }^{2}\lesssim 4.82\times
10^{-15}$ \\ \hline
case 2 & $1.29\times 10^{-3}\lesssim \xi _{\mu \tau }^{2}\lesssim
\allowbreak 4.42\times 10^{-2}$ & $\xi _{e\tau }^{2}\xi _{\mu \tau
}^{2}\lesssim 2.24\times 10^{-16}$ & $\xi _{e\tau }^{2}\lesssim 2.77\times
10^{-14}$ \\ \hline
case 3 & $1.53\times 10^{-3}\lesssim \xi _{\mu \tau }^{2}$ & $\xi _{e\tau
}^{2}\xi _{\mu \tau }^{2}\lesssim 2.24\times 10^{-16}$ & $\xi _{e\tau
}^{2}\lesssim 2.76\times 10^{-14}$ \\ \hline
case 4 & $9.57\times 10^{-4}\lesssim \xi _{\mu \tau }^{2}\lesssim
\allowbreak 1.40\times 10^{-2}$ & $\xi _{e\tau }^{2}\xi _{\mu \tau
}^{2}\lesssim 2.10\times 10^{-17}$ & $\xi _{e\tau }^{2}\lesssim 8.22\times
10^{-15}$ \\ \hline
case\ 5 & $1.02\times 10^{-3}\lesssim \xi _{\mu \tau }^{2}\lesssim
\allowbreak 1.66\times 10^{-2}$ & $\xi _{e\tau }^{2}\xi _{\mu \tau
}^{2}\lesssim 2.93\times 10^{-17}$ & $\xi _{e\tau }^{2}\lesssim 9.65\times
10^{-15}$ \\ \hline\hline
\end{tabular}
\end{center}
\caption{Constraints on the mixing parameters $\protect\xi _{\protect\mu 
\protect\tau }^{2}$,\ $\protect\xi _{e\protect\tau }^{2}\protect\xi _{%
\protect\mu \protect\tau }^{2}\;$and $\protect\xi _{e\protect\tau }^{2}\;$%
for the five cases mentioned in the text. The two former are generated from $%
\Delta a_{\protect\mu }\;$and $\Gamma \left( \protect\mu \rightarrow e%
\protect\gamma \right) \;$respectively, while the latter comes from the
combination of the lower limit on $\protect\xi _{\protect\mu \protect\tau
}^{2}\;$and the upper bound on $\protect\xi _{e\protect\tau }^{2}\protect\xi %
_{\protect\mu \protect\tau }^{2}$.}
\label{tab:mutao}
\end{table}

The first bounds come from $\Delta a_{\mu }$. We use the estimated value of
it,$\;$given by \cite{Maria} at 95\% C.L. eq. (\ref{Maria K}). Since $\Delta
a_{\mu }\;$in eq. (\ref{Maria K}) is positive, lower and upper bounds for
the FC vertex $\xi _{\mu \tau }\;$can be gotten at 95\%\ C.L. The results
are indicated in table (\ref{tab:mutao}) column 1. The lower bounds in each
case are obtained when $m_{A^{0}}\approx 115\;$GeV, and using the minimum
value of $\Delta a_{\mu }$ in eq. (\ref{Maria K}), while the upper bounds
are obtained when $A^{0}\;$is very heavy and using the upper limit for $%
\Delta a_{\mu }\;$in eq.(\ref{Maria K}). From these results a quite general
and conservative allowed interval can be extracted 
\begin{equation}
7.62\times 10^{-4}\lesssim \xi _{\mu \tau }^{2}\lesssim 4.44\times 10^{-2}.
\label{genmutao}
\end{equation}

Furthermore, upper bounds for the product $\xi _{\mu \tau }^{2}\xi _{e\tau
}^{2}$ are obtained from the expression of the decay width $\Gamma \left(
\mu \rightarrow e\gamma \right) $ in eq. (\ref{tamufot}) and from the
experimental upper limit $\Gamma \left( \mu \rightarrow e\gamma \right) \leq
3.6\times 10^{-30}$ GeV\ \cite{data particle}. The most general upper bounds
are obtained for $A^{0}\;$very heavy.$\;$The results are shown in table (\ref
{tab:mutao}) column 2. From this table we infer that quite generally, the
upper limit is 
\begin{equation}
\xi _{e\tau }^{2}\xi _{\mu \tau }^{2}\lesssim 2.24\times 10^{-16}
\end{equation}

Moreover, combining these upper limits with the lower bounds on $\xi _{\mu
\tau }^{2}$ given in the first column of table (\ref{tab:mutao}), we find
upper limits on $\xi _{e\tau }^{2}.$\ The results appear on table (\ref
{tab:mutao}) third column and the general bound can be written as 
\begin{equation}
\xi _{e\tau }^{2}\lesssim 2.77\times 10^{-14}\;.  \label{genemutao}
\end{equation}
Noteworthy, these constraints predict a strong hierarchy between the mixing
elements $\xi _{\mu \tau }\;$and $\xi _{e\tau }\;$i.e. $\left| \xi _{e\tau
}\right| <<\left| \xi _{\mu \tau }\right| $\ and they differ by at least
five orders of magnitude.

\begin{figure}[tbph]
\centerline{\hbox{ \hspace{0.2cm}
    \includegraphics[width=6.5cm]{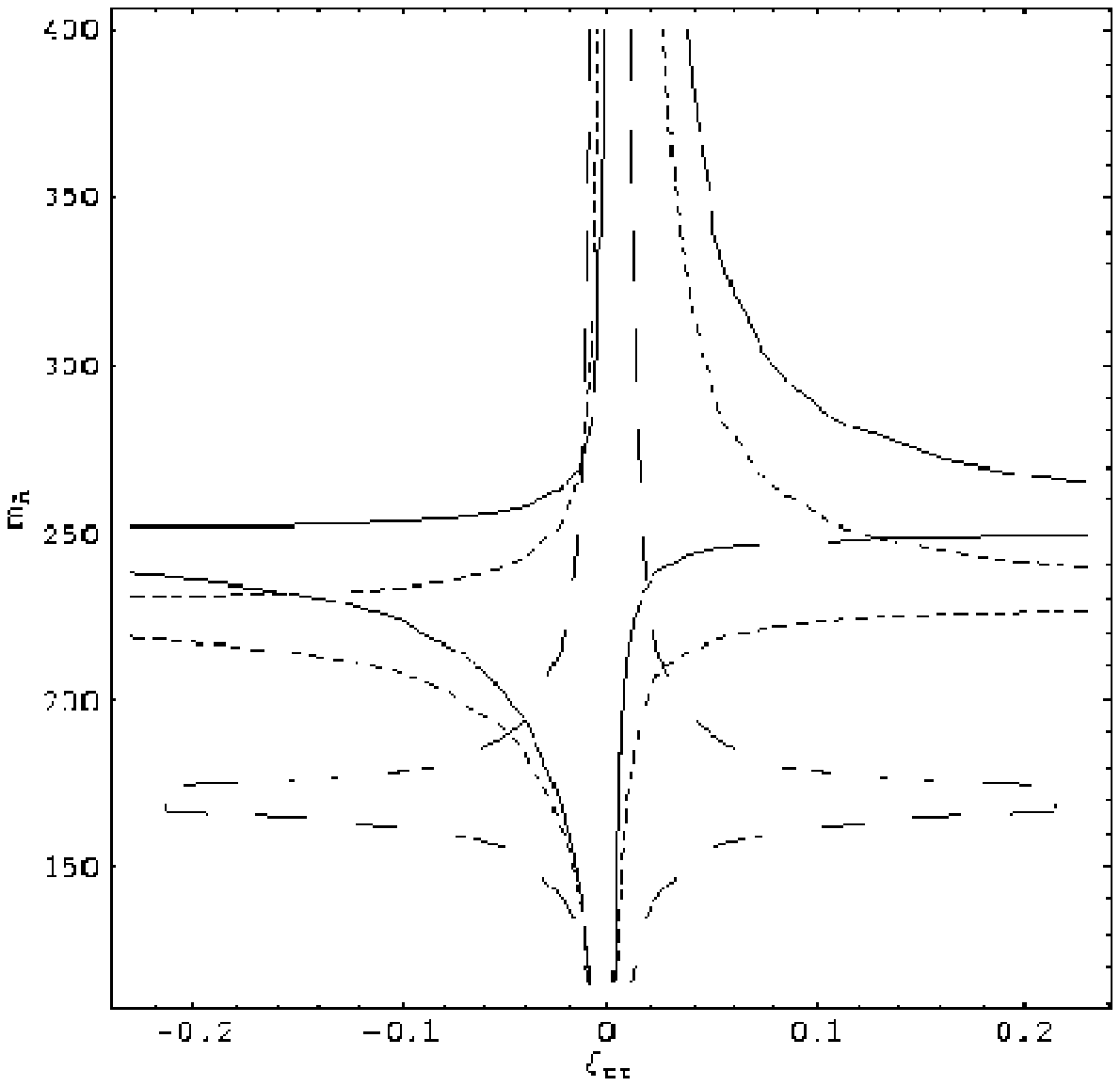}
    \hspace{0.3cm}
    \includegraphics[width=6.5cm]{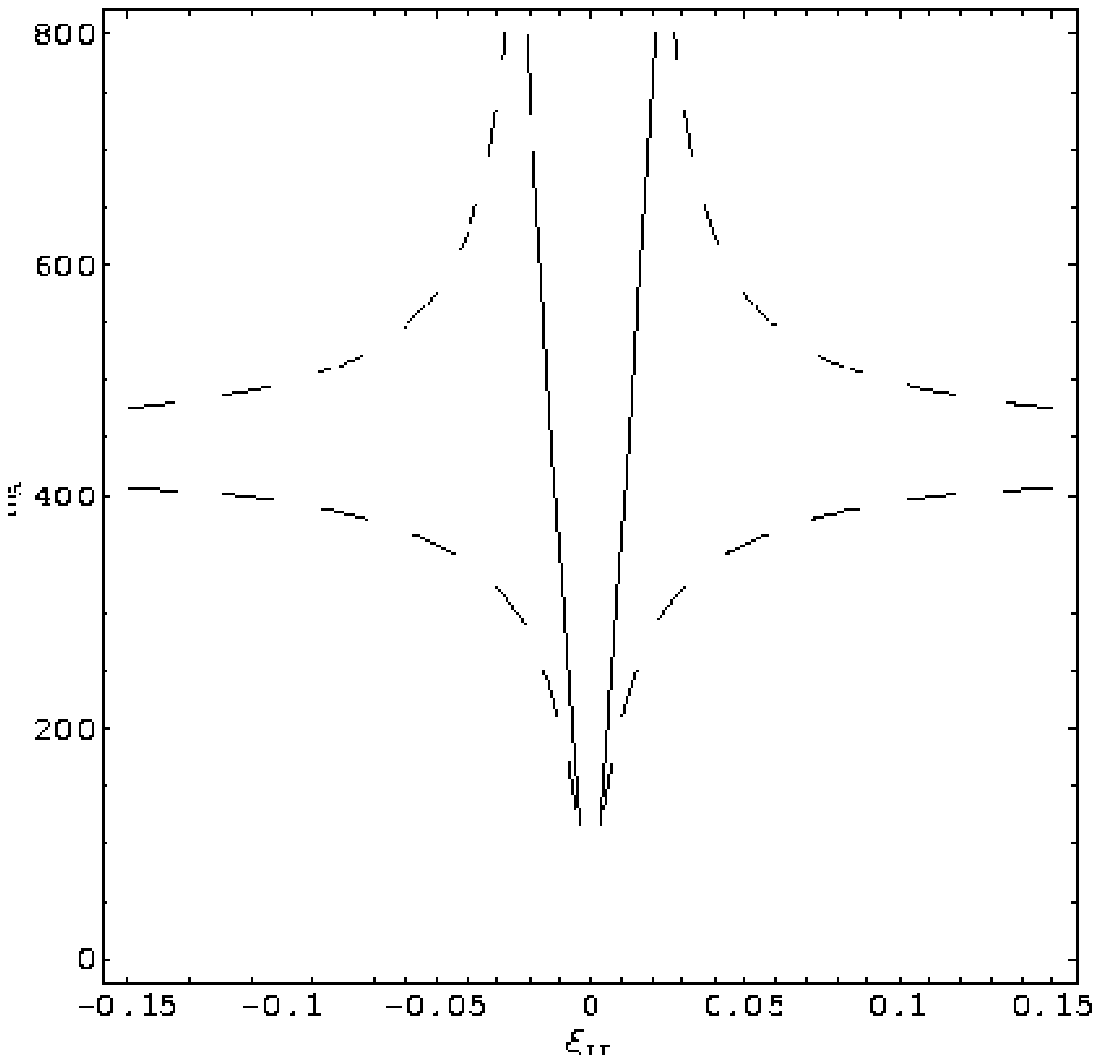}
    }
}
\caption{\textit{Contourplots in the $\protect\xi _{\protect\tau \protect\tau
}-m_{A^{0}}\;$plane for each of the five cases cited in the text. On left:
case 1 corresponds to the long-dashed line, case 4 to the short-dashed line,
and case 5 to the solid line. On right: case 2 corresponds to dashed line,
and case 3 is solid line.}}
\label{fig:taotao}
\end{figure}
\begin{figure}[tbph]
\centerline{\hbox{ \hspace{0.2cm}
    \includegraphics[width=6.5cm]{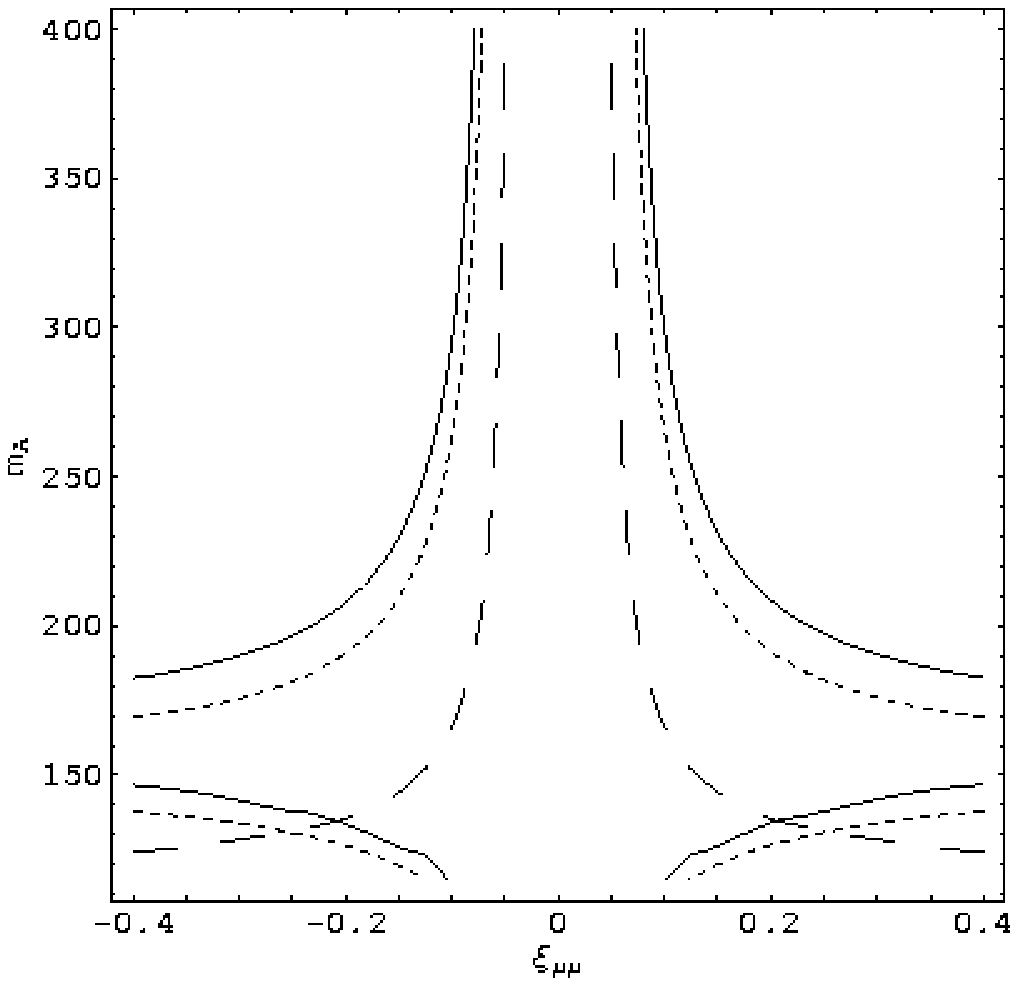}
    \hspace{0.3cm}
    \includegraphics[width=6.5cm]{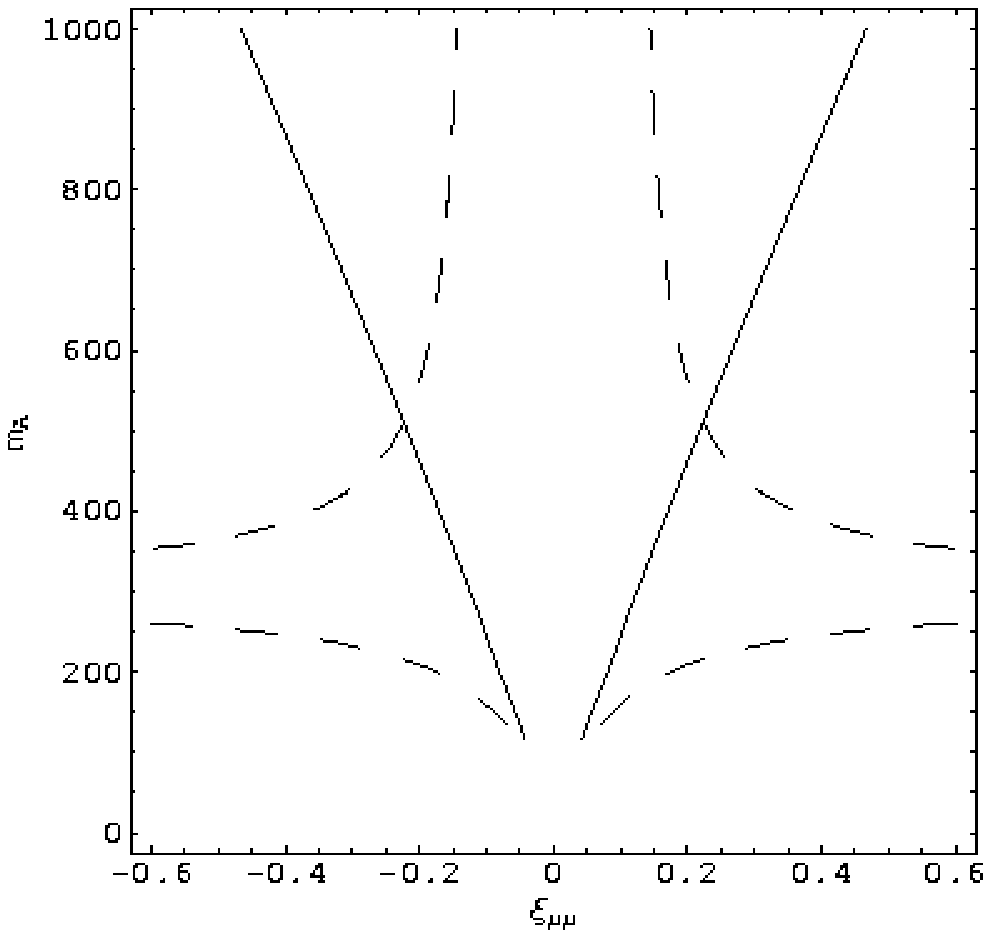}
    }
  }
\caption{\textit{Contourplots in the $\protect\xi _{\protect\mu \protect\mu
}-m_{A^{0}}\;$plane for each of the five cases cited in the text. On left:
case 1 corresponds to the long-dashed line, case 4 to the short-dashed line,
and case 5 to the solid line. On right: case 2 corresponds to dashed line,
and case 3 is solid line.}}
\label{fig:mumu}
\end{figure}

On the other hand, from eq. (\ref{tamufot}) we see that the decay widths $%
\Gamma \left( \tau \rightarrow \mu \gamma \right) $ and $\Gamma \left( \tau
\rightarrow \mu \mu \mu \right) $ depend on two mixing vertices $\xi _{\mu
\tau }^{2},$\ $\xi _{\tau \tau }\;$and $\xi _{\mu \tau }^{2},$ $\xi _{\mu
\mu }$ respectively. Then, we can get conservative constraints on the
diagonal mixing vertices $\xi _{\tau \tau }$, $\xi _{\mu \mu }\;$by using
once again the lower bounds on $\xi _{\mu \tau }^{2}\;$obtained from $\Delta
a_{\mu }$.$\;$Since $\Gamma \left( \tau \rightarrow \mu \gamma \right) $,
and $\Gamma \left( \tau \rightarrow \mu \mu \mu \right) \;$are rather
complicate functions of $\xi _{\tau \tau }$, and $\xi _{\mu \mu }\;$%
respectively, we present these constraints in the form of contourplots in
the $m_{A^{0}}-\xi _{\tau \tau }\;$plane and the $m_{A^{0}}-\xi _{\mu \mu }\;
$plane, figures \ref{fig:taotao} and \ref{fig:mumu}, each one for the five
cases.

We can check that for each contourplot there is a value of $m_{A^{0}}\;$for
which $\xi _{\tau \tau }\;$or$\;\xi _{\mu \mu }\;$stays unconstrained, and
they are shown in tables \ref{tab:taotao}, and \ref{tab:mumu} respectively.
Additionally, the bounds for $\xi _{\tau \tau },\;\xi _{\mu \mu }\;$when $%
m_{A^{0}}$ is very large$\;$and when $m_{A^{0}}\approx 115\;$GeV are also
included in tables \ref{tab:taotao}, \ref{tab:mumu} respectively.$\;$We see
that the general constraints for $\xi _{\tau \tau }$ read 
\begin{eqnarray}
-1.8\times 10^{-2} &\lesssim &\xi _{\tau \tau }\lesssim 2.2\times 10^{-2}\;,
\notag \\
-1.0\times 10^{-2} &\lesssim &\xi _{\tau \tau }\lesssim 1.0\times 10^{-2}
\end{eqnarray}
for $m_{A^{0}}$ very large and for $m_{A^{0}}\approx 115$ GeV respectively.
These constraints are valid for all cases, except for the third case with $%
m_{A^{0}}\;$very large, since in that scenario $\xi _{\tau \tau }\;$remains
unconstrained. Now, for $\xi _{\mu \mu }\;$the general bounds read 
\begin{eqnarray}
\left| \xi _{\mu \mu }\right| &\lesssim &0.12 ,  \notag \\
\left| \xi _{\mu \mu }\right| &\lesssim &0.13
\end{eqnarray}
for $m_{A^{0}}$ very large and for $m_{A^{0}}\approx 115$ GeV respectively.
Once again these constraints are not valid for the third case when $%
m_{A^{0}}\;$is very large, but are valid in all the other cases.

The presence of diagonal mixing vertices could play a crucial role in
looking for FCNC. At this respect, it is worthwhile to point out that the
relative couplings (the quotient between the Yukawa couplings of the new
physics and the SM Yukawa couplings) in model type III 
\begin{eqnarray}
\frac{\left( g_{H^{0}\overline{f}f}\right) }{\left( g_{\phi ^{0}\overline{f}%
f}^{SM}\right) }=\;\cos \alpha  &-&\frac{\sqrt{2}M_{W}\xi _{ff}}{gm_{f}}\sin
\alpha   \notag \\
\frac{\left( g_{h^{0}\overline{f}f}\right) }{\left( g_{\phi ^{0}\overline{f}%
f}^{SM}\right) }=-\sin \alpha  &+&\frac{\sqrt{2}M_{W}\xi _{ff}}{gm_{f}}\cos
\alpha 
\end{eqnarray}
are not universal because of the contribution of the factor $\xi _{ff}/m_{f}$%
. As a manner of example, the relative couplings $H^{0}\tau \overline{\tau }$%
, $H^{0}\mu \overline{\mu }$, $H^{0}e\overline{e}$ are in general different,
a deviation from the universal behavior of the relative couplings of the
type $Hf\overline{f}$ could be a clear signature of FCNC at the tree level.
In the case of the 2HDM type I, all the relative couplings of the form $Hf%
\overline{f}\;$for a certain Higgs are equal. In the model type II, relative
couplings of this form are equal for all up-type fermions, and for all
down-type fermions. By contrast, in the model type III all relative
couplings of the form $Hf\overline{f}$ can be different each other. In
addition, in the 2HDM with FCNC the pseudoscalar Higgs couples to a pair of
fermions of the same flavor by means of the matrix element $\xi _{ff}$,
different from the case of the 2HDM with no FCNC in which the pseudoscalar
couples through the mass of the corresponding fermion. 
\begin{table}[tbp]
\begin{center}
\centering
\vskip0.1 in 
\begin{tabular}{||l||l|l|l||}
\hline\hline
& 
\begin{tabular}{l}
Values of $m_{A^{0}}\;$for \\ 
$\xi _{\tau \tau }\;$unconstrained
\end{tabular}
& 
\begin{tabular}{l}
$\xi _{\tau \tau }\;$intervals for \\ 
$m_{A^{0}}\;$very large
\end{tabular}
& 
\begin{tabular}{l}
$\xi _{\tau \tau }\;$intervals \\ 
for $m_{A^{0}}\approx 115\;$GeV
\end{tabular}
\\ \hline
case 1 & $m_{A^{0}}\approx 170\;$GeV & $-0.0072\lesssim \xi _{\tau \tau
}\lesssim 0.0072$ & $-0.010\lesssim \xi _{\tau \tau }\lesssim 0.010$ \\ 
\hline
case 2 & $m_{A^{0}}\approx 440\;$GeV & $-0.018\lesssim \xi _{\tau \tau
}\lesssim 0.018$ & $-0.0043\lesssim \xi _{\tau \tau }\lesssim 0.0043$ \\ 
\hline
case 3 & ----------------- & unconstrained & $-0.0036\lesssim \xi _{\tau
\tau }\lesssim 0.0036$ \\ \hline
case 4 & $m_{A^{0}}\approx 228\;$GeV & $-0.0075\lesssim \xi _{\tau \tau
}\lesssim 0.022$ & $-0.0094\lesssim \xi _{\tau \tau }\lesssim 0.0036$ \\ 
\hline
case 5 & $m_{A^{0}}\approx 250\;$GeV & $0.00024\lesssim \xi _{\tau \tau
}\lesssim 0.021$ & $-0.0093\lesssim \xi _{\tau \tau }\lesssim 0.0026$ \\ 
\hline\hline
\end{tabular}
\end{center}
\caption{Bounds extracted from the contourplots shown in fig. \ref
{fig:taotao}. The first column indicates the values of $m_{A^{0}}\;$for
which $\protect\xi _{\protect\tau \protect\tau }\;$stays unconstrained for
each of the five cases. The second and third columns show the allowed
intervals on $\protect\xi _{\protect\tau \protect\tau }\;$when $m_{A^{0}}\;$%
is very large and when $m_{A^{0}}\approx 115$ GeV respectively.}
\label{tab:taotao}
\end{table}

\begin{table}[tbp]
\begin{center}
\centering
\vskip0.1 in 
\begin{tabular}{||l||l|l|l||}
\hline\hline
& 
\begin{tabular}{l}
Values of $m_{A^{0}}\;$for \\ 
$\xi _{\mu \mu }\;$unconstrained
\end{tabular}
& 
\begin{tabular}{l}
$\xi _{\mu \mu }\;$intervals for  \\ 
$m_{A^{0}}$ very large
\end{tabular}
& 
\begin{tabular}{l}
$\xi _{\mu \mu }\;$intervals \\ 
for $m_{A^{0}}\approx 115\;$GeV
\end{tabular}
\\ \hline\hline
case 1 & $m_{A^{0}}\approx 115\;$GeV & $\left| \xi _{\mu \mu }\right|
\lesssim 0.043$ & unconstrained \\ \hline
case 2 & $m_{A^{0}}\approx 300\;$GeV & $\left| \xi _{\mu \mu }\right|
\lesssim 0.12$ & $\left| \xi _{\mu \mu }\right| \lesssim 0.055$ \\ \hline
case 3 & -------------------- & unconstrained & $\left| \xi _{\mu \mu
}\right| \lesssim 0.043$ \\ \hline
case 4 & $m_{A^{0}}\approx 152\;$GeV & $\left| \xi _{\mu \mu }\right|
\lesssim 0.058$ & $\left| \xi _{\mu \mu }\right| \lesssim 0.13$ \\ \hline
case 5 & $m_{A^{0}}\approx 163\;$GeV & $\left| \xi _{\mu \mu }\right|
\lesssim 0.061$ & $\left| \xi _{\mu \mu }\right| \lesssim 0.11$ \\ 
\hline\hline
\end{tabular}
\end{center}
\caption{Bounds extracted from the contourplots shown in fig. \ref{fig:mumu}%
. The first column indicates the values of $m_{A^{0}}\;$for which $\protect%
\xi _{\protect\mu \protect\mu }\;$stays unconstrained for each one of the
five cases. The second and third columns show the allowed intervals on $%
\protect\xi _{\protect\mu \protect\mu }\;$when $m_{A^{0}}\;$is very large$\;$%
and when $m_{A^{0}}\approx 115\;$GeV$\;$respectively.}
\label{tab:mumu}
\end{table}

We can check that our bounds on $\xi _{\tau \tau }\;$still permits either a
huge suppression or a huge enhancement on the couplings $H\tau \overline{%
\tau }$\ with $H\;$being any neutral Higgs. For instance, in the case 5 by
using $\xi _{\tau \tau }\simeq 0.01\;$the coupling $h^{0}\tau \overline{\tau 
}\;$vanishes, while using $\xi _{\tau \tau }\simeq 0.1$,$\;$we see that the
contribution coming from the term proportional to $\xi _{\tau \tau },\;$is
about 10 times larger in magnitude than the contribution coming from the
term proportional to the tau mass. In the case 5 both values $\xi _{\tau
\tau }=0.01\;$and\ $\xi _{\tau \tau }=0.1\;$are allowed at least for $%
m_{A^{0}}$\ around $250\;$GeV, as it is shown in fig. (\ref{fig:taotao}).
However, for a very light ($m_{A^{0}}\approx 115\;$GeV), or a very heavy
pseudoscalar Higgs boson, constraints on $\xi _{\tau \tau }\;$are
considerably stronger in all cases, see table \ref{tab:taotao}. Similarly,
strong or weak couplings of$\;H^{0}\tau \tau \;$and/or $A^{0}\tau \overline{%
\tau }\;$are still allowed. Of course, the same pattern is accomplished by
the bounds for $\xi _{\mu \mu }\;$since they are weaker than the constraints
on $\xi _{\tau \tau }$.

Based on the bounds obtained above, we are able to estimate upper limits for
some leptonic decays by using the expressions (\ref{tamufot}),\ (\ref{R}),\
and (\ref{Llll}). We shall assume that $\left| \xi _{\tau \tau }\right|
\approx \left| \xi _{ee}\right| \lesssim 0.1$,$\;$from this assumption we
can obtain the following upper bounds
\begin{eqnarray}
\Gamma \left( \tau \rightarrow e\gamma \right)  &\lesssim &1.5\times
10^{-27}\;,  \notag \\
\Gamma \left( \tau \rightarrow eee\right)  &\lesssim &5\times 10^{-29}.
\end{eqnarray}
If we compare with the current experimental upper bounds $\Gamma \left( \tau
\rightarrow e\gamma \right) \leq 6.12\times 10^{-18}$, $\Gamma \left( \tau
\rightarrow eee\right) \leq 6.57\times 10^{-18}$ \cite{data particle}, we
see that these decays are predicted to be very far from the reach of next
generation experiments in the context of the 2HDM type III, unless that $\xi
_{\tau \tau },$and/or $\xi _{ee}\;$acquire unexpectedly large values.

\section{Concluding remarks}

We calculated some constraints on lepton flavor violating vertices in the
framework of the 2HDM type III, by assuming $m_{h^{0}}\approx 115GeV\;$and $%
m_{A^{0}}\gtrsim m_{h^{0}}$. Since the most recent estimations of $\Delta
a_{\mu },$\ still provides an important window for new Physics, we obtain
from it lower and upper bounds on the mixing vertex $\xi _{\mu \tau }\;$at
95\% C.L. Specifically, an allowed interval of $7.62\times 10^{-4}\lesssim
\xi _{\mu \tau }^{2}\lesssim 4.44\times 10^{-2}\;$was found in a quite wide
region of parameters. Of course, we should realize that both SM test and
experimental measurements of $a_{\mu }\;$are still being scrutinized and
current results are not definitive at all. However, if not severe changes
occur in forthcoming experiments and SM estimations, these constraints could
continue being valid at least at a lower confidence level. Future
improvements on both estimations should elucidate this point.

Based on these constraints, and on the leptonic decays $\mu \rightarrow
e\gamma ,\;\tau \rightarrow \mu \gamma ,\;$and $\tau \rightarrow \mu \mu \mu
\;$we got the following conservative and quite general bounds on LFV
vertices 
\begin{equation}
\xi _{e\tau }^{2}\lesssim 2.77\times 10^{-14}.
\end{equation}
For $\xi _{\tau \tau }$ we find 
\begin{eqnarray}
-1.8\times 10^{-2} &\lesssim &\xi _{\tau \tau }\lesssim 2.2\times 10^{-2}, 
\notag \\
\left| \xi _{\tau \tau }\right|  &\lesssim &1.0\times 10^{-2}
\end{eqnarray}
for $m_{A^{0}}$ very large and for $m_{A^{0}}\approx 115$ GeV respectively.
For $\left| \xi _{\mu \mu }\right| $ we find 
\begin{eqnarray}
\left| \xi _{\mu \mu }\right|  &\lesssim &0.13,  \notag \\
\left| \xi _{\mu \mu }\right|  &\lesssim &0.15
\end{eqnarray}
for $m_{A^{0}}$ very large and for $m_{A^{0}}\approx 115$ GeV respectively.
Notwithstanding, those bounds on $\xi _{\tau \tau }\;$and $\xi _{\mu \mu }\;$%
are considerably relaxed for certain specific values of $m_{A^{0}}$. In that
case, the room available for them is such that they permit either a strong
enhancement or a strong suppression of the couplings $H\tau \overline{\tau }%
\;$and/or $H\mu \overline{\mu }\;$for any neutral Higgs boson. Additionally,
according to these constraints we find that$\;\left| \xi _{\mu \tau }\right|
\;$is at least five orders of magnitude larger than $\left| \xi _{e\tau
}\right| \;$revealing a strong hierarchy between them.

Furthermore, we estimate upper limits on the decay widths $\Gamma (\tau
\rightarrow e\gamma )$, and$\;\Gamma (\tau \rightarrow eee)$, finding that
they are basically hopeless to look for LFV processes in near future
experiments, at least in the framework of the 2HDM type III with heavy Higgs
bosons.



\section{Acknowledgments}

This work was supported by COLCIENCIAS (Colombia), and Banco de la
Rep\'{u}blica (Colombia).

\end{document}